IAC-22, B4,4,13, x69463

# AEROS: Oceanographic Hyperspectral Imaging and Argos-Tracking CubeSat

**Sophie Prendergast[a]\*, Cadence Payne[b], Miles Lifson[b], Christian Haughwout[b], Marcos Tieppo[c], Paulo Figueiredo[d], André Guerra[d], Alexander Costa[d], Helder Magalhães[d], Tiago Hormigo[e], Francisco Câmara[e], Carlos Mano[e], Pedro Pinheiro[e], Alvin D. Harvey[b], Bruno Macena[a], Luis F. Azevedo[f], Miguel Martin[d], Tiago Miranda[c], Eduardo Pereira[c], João Faria[f], Inês Castelão[g], Catarina Cecilio[g], Emanuel Castanho[h], Kerri Cahoy[b], Manuel Coutinho[i], Helder Silva[i], Jorge Fontes[a]**

[a] *Institute of Marine Sciences - Okeanos, University of the Azores, Rua Professor Doutor Frederico Machado 4, 9901-862 Horta, Portugal, okeanos.secretariado@uac.pt*
[b] *Massachusetts Institute of Technology, 77 Massachusetts Avenue, Cambridge, MA 02139, USA, kcahoy@mit.edu*
[c] *Institute for Sustainability and Innovation in Structural Engineering (ISISE), University of Minho, Azurém Campus, 4800-058, Guimarães, Portugal, isise@civil.uminho.pt*
[d] *CEiiA, Av. Dom Afonso Henriques, 1825. 4450-017 Matosinhos, Portugal, aeros@ceiia.com*
[e] *Spin.Works S.A., Rua de Fundões n.º151, 3700-121 São João da Madeira, Portugal, info@spinworks.pt*
[f] *DSTelecom, Rua de Pintancinhos s/n, 4700-727 Braga, Portugal, geral@dstelecom.pt*
[g] *CoLAB +ATLANTIC, Estrada da Malveira da Serra 920, 2750-834 Cascais, Portugal, info@colabatlantic.com*
[h] *AD AIR Centre, Terinov-Canada de Belém S/N, Terra Chã, 9700-702, Angra do Heroísmo, Ilha Terceira, Portugal, info@aircentre.org*
[i] *EDISOFT S.A., Rua Calvet Magalhães 245, 2770-153 Paço de Arcos, Portugal, info@edisoft.pt*
\* Corresponding Author

**Abstract**

AEROS is a 3U CubeSat pathfinder toward a future ocean-observing constellation, which will operate in a 500 km Sun-Synchronous orbit, targeting the Portuguese Atlantic region. AEROS features a miniaturized, high-resolution Hyperspectral Imager (HSI), a 5MP RGB camera, and a Software Defined Radio (SDR) to interface with Argos, a globally distributed system of remote platforms that collect and relay oceanographic and meteorological data. These sensors will facilitate the advancement of Portuguese scientific and technological knowledge, and international space industry partnerships, with data processed and aggregated for end-users in a new web-based Data Analysis Center (DAC).

The HSI has 150 spectrally contiguous bands covering visible to near-infrared (470 nm – 900 nm) with 10 nm bandwidth (full width at half maximum). The HSI collects ocean color data to support studies of oceanographic characteristics (e.g. primary productivity, mesoscale ocean fronts, and eddies) known to influence the spatio-temporal distribution and movement behavior of marine organisms. To inform performance estimates for the HSI, radiometric analyses were conducted to characterize band sensitivity given varying atmospheric and target conditions. Simulated HSI images were generated to assess suitability for scientific purposes through interpolation of spectral information from Sentinel-2 imagery combined with the AEROS's sensor spectral responses (real calibrated data) and mission-specific parameters.

Usage of an SDR expands AEROS's operational and communication range and allows for remote reconfiguration. The SDR receives, demodulates, and retransmits short duration messages (401.650 MHz + 30kHz), from Argos sources including tagged marine organisms, vessels, autonomous vehicles, subsurface floats, and buoys. This allows AEROS to retransmit messages from Argos platforms to ground and processing stations that compute platform locations using Doppler effect measurements.

The future DAC will collect, store, process, and analyze acquired data, taking advantage of its ability to disseminate data across the stakeholders and the scientific network. Correlation of animal-borne Argos platform locations and oceanographic data will advance fisheries management, ecosystem-based management, monitoring of marine protected areas, and bio-oceanographic research in the face of a rapidly changing environment. For example, correlation of oceanographic data collected by the HSI, geolocated with supplementary images from the RGB camera and fish locations, will provide researchers with near real-time estimates of essential oceanographic variables within areas selected by species of interest (e.g., sharks, rays, and tuna). This is essential for determining how keystone species, or those vulnerable to overexploitation, interact with their environment. Other relevant applications include monitoring sediment transport, erosion, pollution, and human activities.

**Acronyms/Abbreviations**

Colored Dissolved Organic Matter (CDOM)
Commercial-Off-The-Shelf (COTS)
Complementary Metal-Oxide Semiconductor (CMOS)
Data Analysis Center (DAC)
Digital Signals Processing (DSP)
Earth Observation (EO)
Essential Ocean Variables (EOV)
Extract, Transform and Load (ETL)
GNU Radio Companion (GRC)
Harmful Algal Bloom (HAB)
Hyperspectral Imager (HSI)
Infrared (IR)
Marine Protected Areas (MPAs)
Red, Blue, Green (RBG)
Region of Interest (ROI)
Sea Surface Salinity (SSS)
Sea Surface Temperature (SST)
Signal-to-Noise Ratio (SNR)
Software Defined Radio (SDR)

## 1. Introduction

### 1.1. Motivation

Since the onset of the industrial period the world's oceans have acted as a biogeochemical buffer to the negative impacts of anthropogenic activities on Earth's climate and resources. As the world's largest carbon sink, global oceans have absorbed approximately 30% of anthropogenic $CO_2$ emissions [1] while also capturing 90% of the excess heat produced by the warming planet since 1971 [2, 3]. However, this has resulted in significant impact to marine biophysical systems ranging from changing ocean temperature and ocean acidification to expansion of oxygen minimum zones [4]. This in turn adds additional pressure to marine ecosystems and resources that are already strained from human activities such as overfishing, pollution and resource extraction [5].

Remote sensing satellite systems are a vital tool for oceanographic monitoring as they provide broad coverage and high temporal resolution of Essential Ocean Variables (EOV's). EOV's are known to influence marine systems (i.e., Sea-Surface Temperature (SST), Sea-Surface Salinity (SSS), and ocean color (a proxy for primary productivity)) [6]. The introduction of low-cost CubeSats (nanosatellites) has significantly improved the accessibility and capabilities of Earth Observation (EO) missions carried out by traditional satellite systems [7, 8]. CubeSat platforms host sensors that fill spectral resolution and temporal gaps of pre-existing constellations to enhance coverage within specific Regions of Interest (ROIs) [9]. These features make CubeSats ideal tools for ecologists and conservationists working to monitor marine ecosystem structure, composition, and the impact of human activities within high priority regions (i.e., Marine Protected Areas (MPAs)).

Although there are hundreds of small satellites currently in orbit, very few preform ocean monitoring leaving significant gaps in coverage of scientifically critical measurements and ROI's [10]. The AEROS project aims to address this issue through the development of a CubeSat constellation focused on the exclusive economic zones of the Portuguese Atlantic Region which includes the mainland, Azores, Madeira as well as the extended Portuguese continental shelf (Fig. 1).

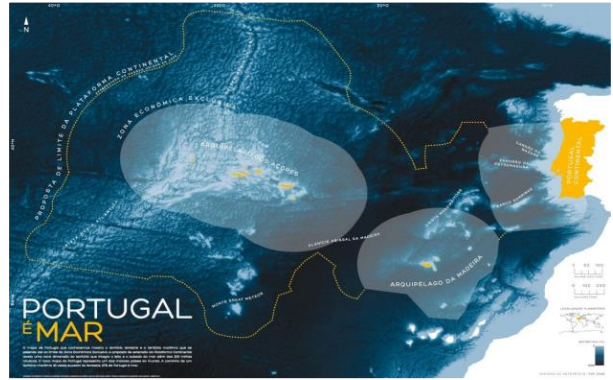

Fig. 1. The AEROS ROI, including the Portuguese Exclusive Economic Zones (translucent blue) and Portuguese Extended Continental Shelf (yellow) image produced by emepc (https://www.emepc.pt/)

### 1.2. The AEROS Mission

The AEROS mission is a 3U (10 x 10 x 30 cm$^3$) CubeSat under development as a precursor for an ocean monitoring constellation that will target the Portuguese Atlantic Region. The mission aims to advance Portuguese know-how regarding technologies that bridge the interactions between ocean and space. AEROS' objectives are as follows: 1) develop and launch a novel CubeSat platform for ocean monitoring, 2) demonstrate miniaturized and efficient high-spectral imaging for forecasting ocean fronts and collect location data from marine fauna fitted with ARGOS and/or LoRaWan linked satellite tags 3) implement data science techniques for classifying and forecasting oceanic evolution to generate additional value-added data for stakeholders, 4) develop flexible software-defined communication modules to support connectivity and network operations of autonomous vehicles and biologging tagging technology (e.g. tagged migratory marine organisms), and 5) establish a Data Analysis Centre (DAC) to collect, process, and analyse data acquired by the AEROS payload (Fig. 2).

To achieve these objectives, AEROS will host three payloads: a 150-band optical and near IR Hyperspectral Imager (HSI); a wide field-of-view (FOV), red-green-blue (RGB) camera; and a Software Defined

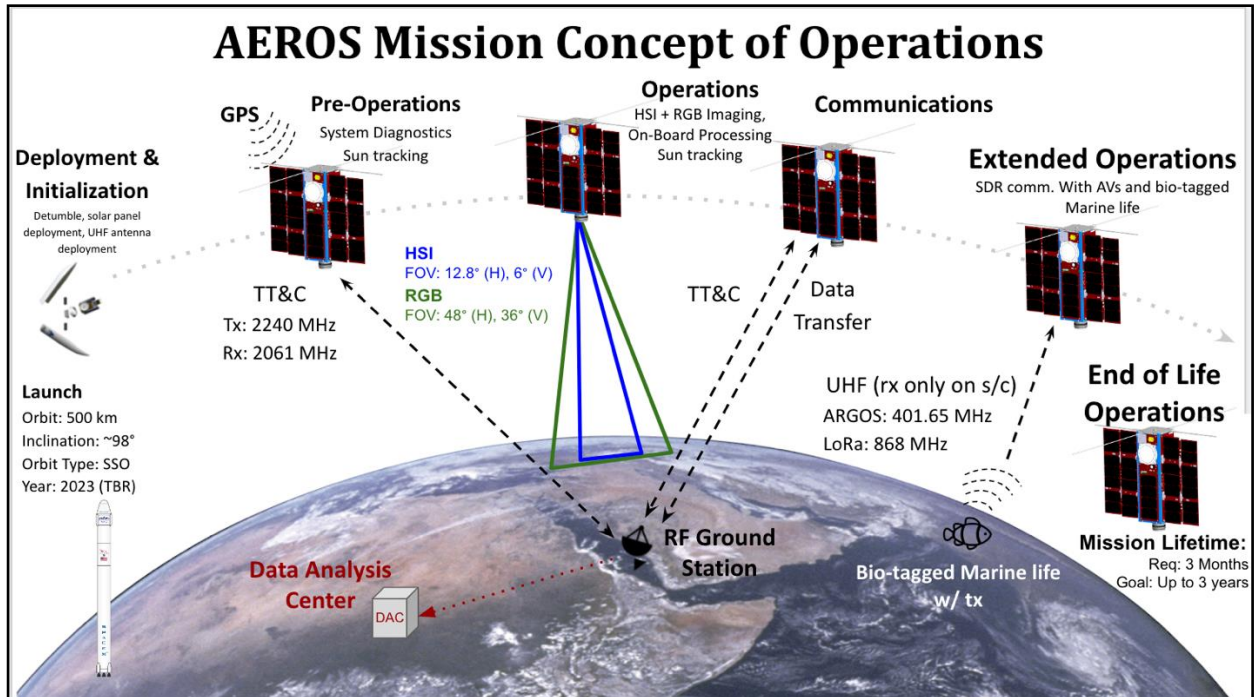

Fig. 2. AEROS Mission Concept of Operation

Radio (SDR) capable of demodulating and retransmitting signals from in-situ ocean assets (i.e., animal-born satellite tags, autonomous vehicles, and buoys) via Argos and LoRaWAN systems. The mission will be placed in an approximately circular 500 km mid-morning, Sun-Synchronous orbit. The AEROS payloads were selected to address the needs of end users across multiple fields of study with three main use cases: (1) understanding the effect of biological EOV's (i.e., phytoplankton/concentration of chlorophyll-a) on the distribution of migratory marine organisms, (2) supporting fisheries and aquaculture management, and (3) contributing to ecosystem-based management strategies and MPA monitoring.

## 2. Technical Development and Approach

AEROS' primary payload is a low-power (5W max), compact (70.8 mm x 70.8 mm x 105 mm$^3$) HSI (Fig. 2). It uses a static spectral filter integrated on top of a Commercial-Off-The-Shelf (COTS) Complementary Metal-Oxide Semiconductor (CMOS) detector to achieve 150 visible and near infrared measurement bands from 470 nm to 900 nm, each with 10 nm bandwidth. A CrystalSpace RBG imager will provide contextual imagery of overlapping ground scenes for the HSI. The HSI will feature a 12.8° by 6° FOV, while the RGB camera will feature a 44.3° by 34° FOV. AEROS also hosts a SDR configured based on a Zynq-7000 system-on-a-chip and uses the GNU Radio software. The SDR supports AEROS' objective to improve flexible connectivity with autonomous vehicles and biologging tagged marine life (Fig. 3).

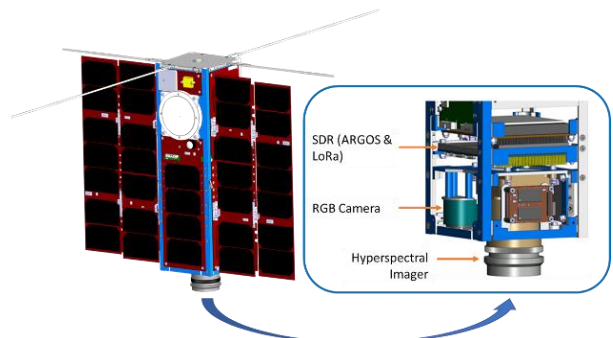

Fig. 3. A CAD rendering of the AEROS CubeSat showing locations of the primary payloads: the SDR, RGB Camera and HIS.

### 2.1. Concept of Operations & Spacecraft Orbit

The AEROS spacecraft will be placed into a near-circular, mid-morning (10:30+60min local time of the descending node) Sun-synchronous orbit at approximately 500 km of altitude as a ride-share. The mission does not feature propulsion and will slowly decay in altitude, meaning that the sun-synchronous properties of the orbit will not be preserved. Nevertheless, this orbit still provides coverage of the ROI and reasonable consistency for power generation and lighting conditions for imaging. More information about the spacecraft's lifetime, power generation, and coverage parameters is available in previously published documents [9].

Once the spacecraft is deployed from the launch vehicle, an initial system health and safety check out is performed. AEROS' primary data products are collected during the operations phase and transmitted down to ground stations in Portugal. The data is then transferred to, and processed at, the DAC. Extended operations include the demonstration of SDR connectivity with bio-tagged marine life and autonomous vehicles. In the Iberian Peninsula a spacecraft has an operational lifetime of at least three months.

*2.2. Hyperspectral Imager*

The optical design of the HSI is based on a 50 mm F2.8 COTS lens (see imager parameters in Table 1). Extensive optical characterization was performed including the determination of the Modulation Transfer Function, Point Spread Function, distortion map, straylight response and spectral transmission. Venting holes were added to prevent trapped air and condensation, and detailed structural and thermal Finite Element Method analyses were performed. Last, the assembly underwent vibrational and Thermal Vacuum Chamber tests to ensure imager survivability given storage and operational requirements in the spacecraft's anticipated environmental conditions. The HSI total mass is less than 500g and can operate between -30ºC and 60ºC.

Table 1. HSI Parameters (imaging component)

| Parameter | Value |
|---|---|
| Package Mass | 275 g |
| Package Volume | 86.25 mm x 66 mm x 55 mm |
| Optics Aperture Diameter | ~18 mm |
| Focal Length | 50 mm |
| f/# | 2.8 |
| FOV | 12.8° (full angle, horizontal), 6.0° (vertical), 14.6° (diagonal) |
| Spectral Bandpass ($\lambda max - \lambda min$) | 470 nm - 900 nm |
| Spectral Bandwidth ($\lambda_0 - \lambda_f$) | 10 nm (collimated) |
| Number of Bands | 150 |
| Sensor Type | Line scanner, pushbroom |

The HSI (Fig. 4) contains an Imec LS150 hyperspectral sensor built on the AMS CMV2000, a CMOS imaging sensor wafer. A custom designed stair-case prism is integrated on top of a monochromatic CMOS sensor that has a spectral response of 470 nm - 900 nm. The height of the prism is carefully controlled to produce a sequence of narrow-band filters with increasing center-wavelengths, capturing different wavelengths along contiguous lines of the sensor.

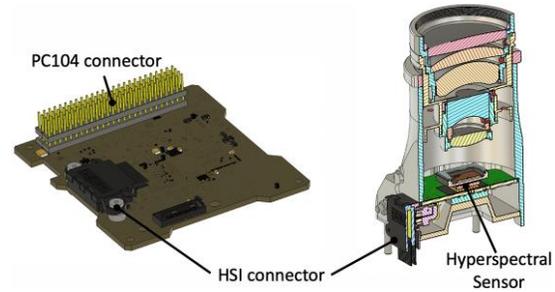

Fig. 4. HSI processing board (left) and cross-sectional view of the HSI and imaging component (right).

The LS150 sensor is a line scan filter, meaning that the present 192 raw spectral bands are disposed along the vertical direction of the sensor, with each filter occupying 5 complete lines. The filters are distributed over two separate active areas. One has 64 filters active in the visible range (470 nm - 600 nm) and the other 128 filters are in the near infrared (600 nm - 900 nm). The active areas are separated by an empty interface zone of 120 rows.

*2.3. RGB Camera*

The RGB camera will allow the AEROS mission a synoptic view of ocean features acquired by the HSI. Although the HSI will enable the observation of oceanography processes in high spatial resolution, its coverage is limited, namely due to constraints imposed by limited CubeSat volume. Thus, the RGB camera was included to improve the coverage of the AEROS dataset and provide larger-scale contextual measurements for the HIS via imaging overlapping ground scenes (Fig. 5). Although some oceanography processes have specific spectral signatures that are not visible to the human eye, in general, RGB images can show features such as river flow, continental shelf currents, and sediment plumes.

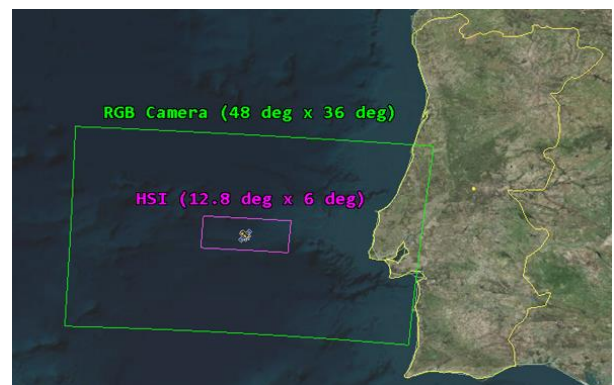

Fig. 5. Image showing the FOV of the RGB Camera in relation to the HSI.

Furthermore, the RGB camera will be used as a "view-finder" for HSI tasking, as well as to reduce spacecraft attitude estimation errors and improve georeferencing and registration of HSI imagery.

*2.4. Software Defined Radio*

AEROS' communication system is an SDR, a system where components that have been traditionally implemented in analog hardware are, instead, implemented in software. While many high-performance SDR solutions are available, a space-qualified, COTS radio was used. The selected model is the TOTEM SDR provided by AlenSpace.

Collecting global observations is a primary mission goal of AEROS; therefore, we intended to acquire signals from globally distributed ARGOS platforms (placed on marine animals), autonomous vehicles, and ocean-surveying buoys. This provides end users with simultaneous position and environmental data, enabling scientists to investigate environmental drivers of animal behavior and enhance oceanographic forecasting. These devices are powered by batteries or solar energy and upload short duration messages (less than one second) at a bit rate of 400 bps on a transmission frequency of 401.65 Mhz ± 30 kHz, characteristics compatible with the AEROS SDR.

The same device is used to acquire LoRa signals. LoRa is the physical layer of LoRaWAN, a software communication protocol that forms a network composed of end devices, gateways, processing servers and end user data. The end devices acquire data from battery-powered sensors and transmit them to the nearest gateway. Transmitted data is then sent to the processing servers where it is prepared for end users. Our SDR is intended to serve as a gateway in space, specifically targeting signal acquisition from low-crowded areas.

*2.5. Data Analysis Center*

In addition to the AEROS spacecraft, the team is developing a DAC. Satellite-acquired data will get downlinked to the Ground Control Station where it will then get passed to the DAC for processing and formatting, prior to being made available to end users.

The DAC collects, stores, processes, and analyzes acquired data by the payload aboard the satellite, as well as data from ground-linked systems (e.g., by integrating ocean forecasts from operational ocean models). The DAC will provide data products with different maturity and configuration levels for multiple forms of data visualization. For the DAC to achieve this, the team must first identify and map data product availability to users' needs.

Though currently early-stage, the DAC architecture (Fig. 6) has three development phases: the back-end; the Extract, Transform and Load (ETL) segment (data intelligence); and the front-end (Fig. 6). The back-end supports data acquisition from diverse sources with varying formats and types. It is also responsible for managing all the acquisition processes, whether on raw data or products with higher maturity levels provided from partners and stakeholders.

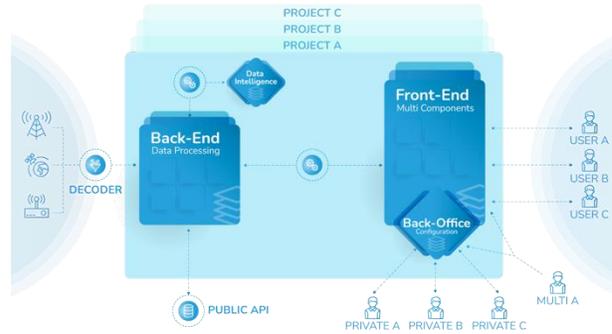

Fig. 6. Data Analysis Center Architecture. Front-end, ETL (data intelligence) and Front-End developed in a micro services approach.

After data acquisition, the ETL phase cleans, harmonizes, and makes data available to users. Availability is either for front-end direct display or for partners and stakeholders to run models, calibrate sensors, etc., inevitably increasing product maturity levels and data product use cases.

Finally, the front-end will display the different data types on a dashboard for users to access. Besides displaying the different product levels, the dashboard will allow data correlation between products from the AEROS constellation, as well as with other data types provided by additional sources, with different maturity levels collected over varying time scales.

*2.6. Future Constellation Coverage*

The AEROS satellite is a precursor for a future ocean observation constellation. To estimate preliminary improvements to revisit times from a larger constellation, a simulation was performed for a uniform 16 satellite constellation featuring four planes of four satellites each at 500 km. The satellites are maintained in a frozen periodic sun-synchronous Flower Constellation with parameters $N_o = 4$, $N_{so} = 4$, $N_c = 1$. Satellites were propagated with a degree 10, order 0 geopotential model. Atmospheric drag, tidal effects, solar radiation pressure, and third body effects were disabled based on an assumption propulsion would be used to counteract these perturbations.

Coverage was simulated for a 30-day period beginning 1 March 2023 with a 0.5-degree coverage grid discretization across the ROI. Coverage was subject to a ground-illumination constraint and assumes both nadir-pointing and that the HSI FOV remains the same in the follow-on constellation.

Coverage was assessed based on a coverage gap duration figure of merit. An ideal constellation would feature a 100% revisit rate after 0 days, corresponding to the top left corner of the graph.

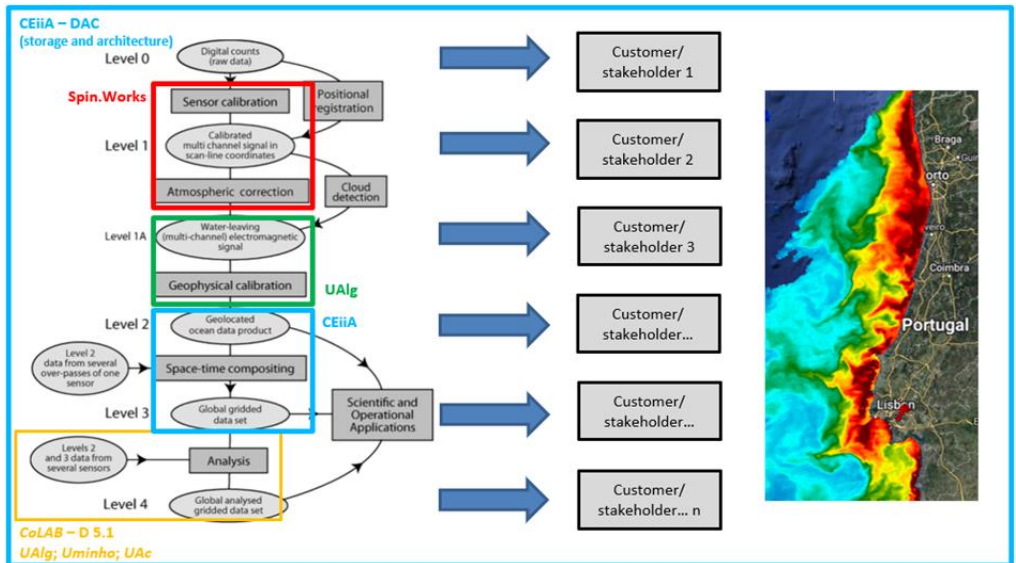

Fig. 7. Different maturity levels following development of the Data Analysis Centre

As the number of satellites increases so does the coverage of the constellation (Fig. **8**). Satellites in each plane are numbered from 0 to 3, with the planes also numbers from 0 to 3. The staggered satellite simulation features satellites (0,0), (1,1), (2,2), (3,3), while the staggered satellite, offset 2 simulation features satellites (0,0), (1,2), (2, 0), (3,2). Further study of various constellation designs and associated concepts of operation will help inform the design of the future constellation.

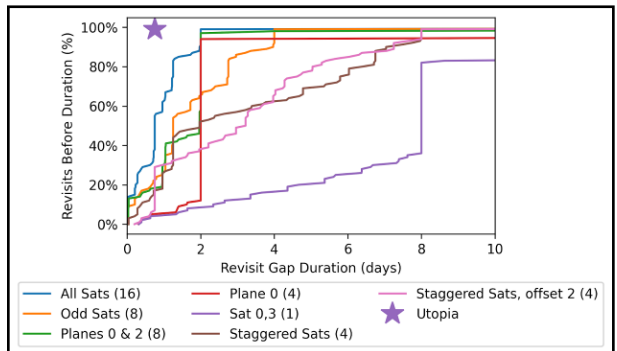

Fig. 8. Follow-on Constellation Coverage Gap Duration. Values in parenthesis refer to the modelled number of satellites

## 3. Expected Data Products and Distribution

The DAC products will be presented in 6 layers (Fig. 8). Level 0 corresponds to the raw data and other information from the sensor/platform/satellite. In this layer users will have access to both platform and sensor data. Level 1 corresponds to the products obtained after the calibration of the instruments/sensors, as well as the geographic and temporal positioning and atmospheric correction. Level 1A, in turn, corresponds to the data after range and geophysical characteristics corrections. In this level some specific products will be expected (e.g., salinity). Conversely, level 2 corresponds to products for scientific and operational applications. This level considers data acquisition during several passes of the sensor through the same area of interest and associated products. Level 3 considers a composition in the space-time domain, presenting as products Global Gridded data sets and time series for spatial compositions. Finally, the last level (Level 4), also corresponds to products for scientific and operational applications. However, this level considers this information (data) with and when correlated with other data sources.

The following section describe how each payload will contribute to the full picture that will be presented in the final DAC dashboard**.**

### 3.1. HSI Use Case

The HSI points nadir and operates as a pushbroom imager. The sensor must be mounted with lines perpendicular to the flight direction, such that the same patch of the Earth surface is observed by all hyperspectral bands at subsequent time periods. The image capture rate is calibrated to expose the sensor at the maximum time required for the satellite to travel the equivalent of one sensor line. Consecutive sensor frames must represent a vertical translation of 5 pixels. At an orbital altitude of 500 km, the ground speed in the image plane is about 7.1 km/s, the Ground Sampling Distance is 55 m, maximum exposure time is 7.8ms (required to maximize Signal-to-Noise Ratio (SNR)), and frames are acquired at about 25.7 frames per second. The final hyperspectral product has a resolution of 112 km x 60 km with 150 calibrated bands.

*3.1.1. HSI Testing Setup*

The HSI has been designed and the imaging package is assembled. An engineering model has been mounted on a sliding rail test bench (Fig. 9) for performance verification testing of the imager's sensor, as well as testing the image processing techniques required to reconstruct the hyperspectral cube.

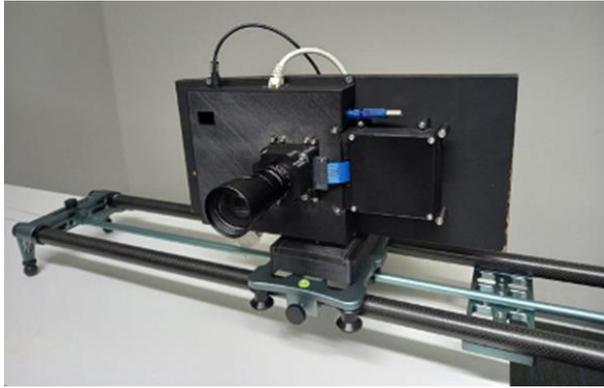

Fig. 9. HSI engineering model mounted on a sliding rail testing bench for performance testing.

*3.1.2. Hypercube Construction in Push-Broom Emulation*

Due to the push-broom operation, hypercubes are constructed by overlapping subsequent hyperspectral frames, each collecting new spectral information from the same ground scene. Actual satellite dynamics introduce shifts between frames (jitter), which drastically affect the final hyperspectral product. Pushbroom operations were emulated via the aforementioned sliding rail to mimic data collection along the spacecraft's flight direction. The subject testing setup velocity was tuned to match the 150 px/s in the image plane, while introducing some unpredictable jitter (noise) to each captured frame that resulted from mechanical perturbation sources (rail vibrations).

For jitter correction, the 1D cross-correlation was first computed between the same patch of terrain at different timestamps. Subsequently, the displacement estimate was refined by performing a quadratic correlation. In the end, the *i+1* image frame is corrected by convolving the entire image frame, using Lanczos Interpolation, such that all bands are subject to the same noise level at every iteration. Jitter effects were clearly diminished along the horizontal axis of the image plane and cumulative displacement (low-frequency movement perturbations) was significantly reduced (Fig. 10). Analyzing the frequency domain, the jitter correction can be defined as a full-spectrum filter that ultimately dampens the jitter signal (Fig. 11).

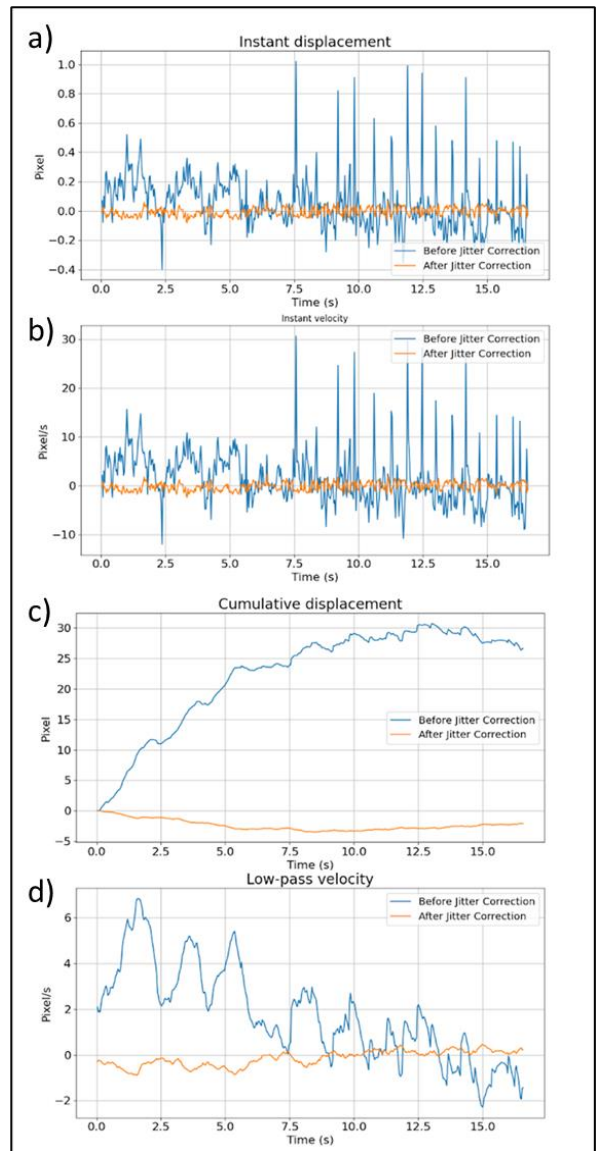

Fig. 10. Comparison of four quantities; a) instant displacement, b) instant velocity, c) cumulative displacement and d) low-pass velocity, along the time domain before and after jitter correction.

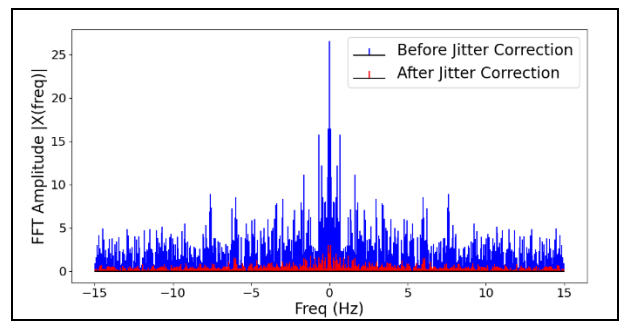

Fig. 11. Frequency domain of the instant displacement before and after jitter correction.

## 3.2. Software Defined Radio Development

As introduced in section 2.4 the SDR is programmed to have two communication protocols: ARGOS and LoRa. The Digital Signal Processing (DSP) software for the ARGOS protocol is made on GnuRadio Companion (GRC), which is an open-source, low-code (only variables within the blocks of the processing chain are manipulated) software platform. Requirements traceability is a critical reference for developing this complex demodulation scheme. These references better inform our understanding of system needs and required methods for validation.

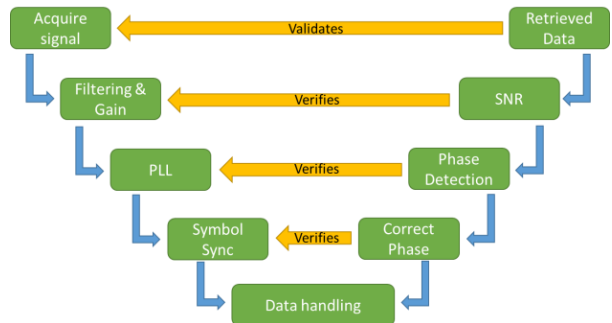

Fig. 12. Steps for acquiring ARGOS signal within the SDR.

Five steps have been identified to acquire the data conveyed in the carrier wave (Fig. 12). To extract the bits modulated on the carrier the signal must reach our satellite antenna, supplied by ISISpace, which is designed to a resonating frequency of 401.65 MHz to match the protocol. The Alenspace UHF front-end will amplify and filter to maximize the SNR. These processes happen first in handwear and are then repeated in software via the GRC code. A Phase Locked Loop (PLL) is then used for phase detection. Lastly, symbol synchronisation is used to achieve an accurate phase input signal. When the DSP sequence is correctly implemented, the GRC output will have identical phases to the input signal. The laboratory test setup includes an ARTIC R2 chip (from an Arribada Horizon ARTIC R2 Developer's kit) that transmits data with the ARGOS modulation scheme for receiving the signal (Fig. 13).

The preliminary results of wired testing are promising, achieving ten encountered preambles in ten transmissions. Receipt of the ARGOS preambles indicates that the system correctly detects all 24 successive bits, validating that the system is outputting phases equal to the inputs. The data is encoded as Manchester code, meaning the phase values are ± 1.1 radians (Fig. 14). The next step involves estimating the anticipated noise interference at our expected orbital placement to further validate system operations.

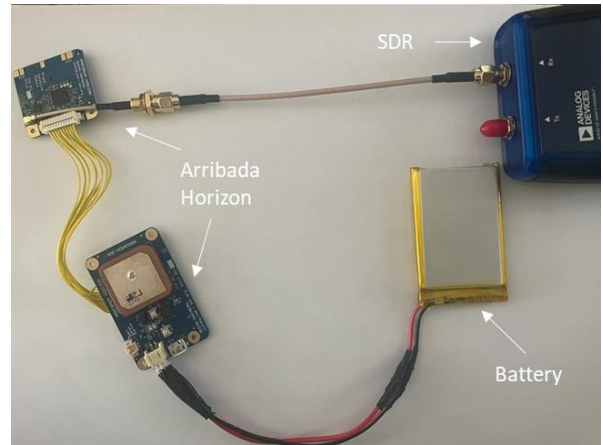

Fig. 13. SDR laboratory setup used for validating system operations with the ARGOS protocol.

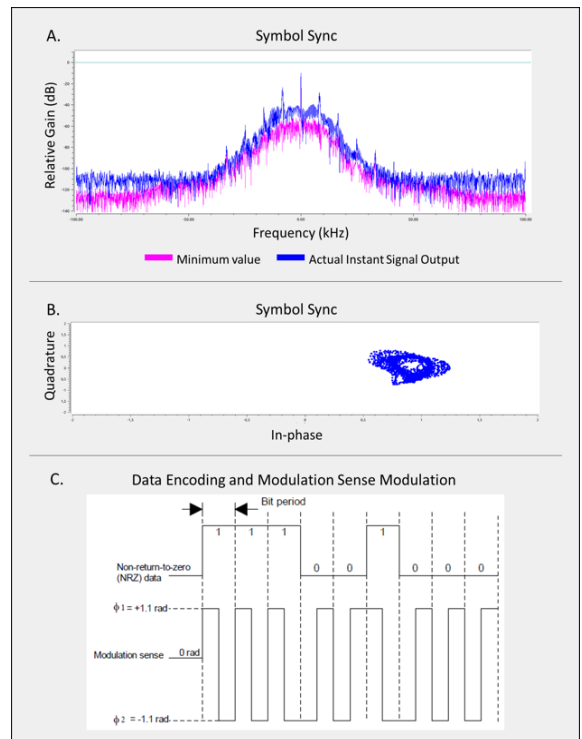

Fig. 14. A) Flowgraph of FFT of received signal (blue) in relation to the minimum value (pink) at the end of the DSP processing chain, B) constellation diagram illustrating the constellation points (blue dots) as expected for Manchester code, C) description of the Manchester Code.

## 3.3. DAC Development

The DAC is in early-stage development. The fully developed system will provide six layers of data products (see Fig. 14). Level 0 corresponds to raw data products provided by sensors, platforms, and satellites. This layer allows users access to high-level platform telemetry, as well as sensor data. Level 1 corresponds to the data products obtained after the instrument/sensors are calibrated, and once the geographic/ temporal positioning and atmospheric correction processes are applied. Level 1A corresponds to the resulting products after range and geophysical characteristic corrections are applied. SSS is one example of an expected Level 1A data product. Level 2 corresponds to products used for scientific and operational applications. These data products consider acquisition of the same product type over multiple passes of the same geographical region. Level 3 considers composition in the space-time domain, providing products such as Global Gridded data sets and time series for spatial compositions. Finally, Level 4 provides data products that are correlated with external data sources.

Currently, a macro functionalities approach is used to assess appropriate data types, users, and methodologies. This approach requires the implementation of a test plan at the end of the DAC development process to ensure efficient use of the system. The approach's first stage considers the execution of unit, functional "black box," and regression tests to ensure continuous operation and integration of all planned tests into the DAC.

DAC development considers the use of agile methodologies for the AEROS project. This approach reinforces code production velocity and ensures that newly added methodologies have no impact on previously developed and tested functionalities leading to potential issues with regression. Reinforcing test executions enables higher execution times in response to changes in the code. Additionally, continuous integration of user feedback is mandatory for the DAC since it allows for the reduction of several issues like version errors, developers' exits, network errors, etc. Continuous integration also allows for near-immediate bug fixing, avoiding versions that are unsuccessful. Tenants, features, roles, users, and permission areas are tested and developed for the DAC, and AEROS foundations tests are scheduled to be performed for APIs and databases.

## 4. Description of AEROS Use Cases

### 4.1. Essential Ocean Variables and Distribution of Marine Megafauna

Satellite data has become essential for understanding the distribution and movement patterns of highly mobile marine species in a changing environment. This is particularly true for species such as pelagic elasmobranchs (i.e., sharks and rays) that face additional pressure from human activities such as commercial fisheries. Satellite tracking systems, such as ARGOS, have allowed researchers to monitor the movements of vulnerable shark and ray species over extensive, and often inhospitable, habitats largely inaccessible to humans [11]. Combined assessments of this information with data from EO missions enables researchers to investigate the environmental factors and mechanisms that drive the observed movement patterns. As a result, researchers have identified biotic and abiotic EOVs such as SST, dissolved oxygen (DO), SSS, ocean color (a proxy for primary productivity), mesoscale eddies, and oceanographic fronts that influence the 3-dimensional distribution of pelagic elasmobranchs [11-13]. However, environmental data from EO missions still lack the required spatial resolution for ground-site or organism-level analyses.

Once fully established, the AEROS constellation will significantly increase the temporal accuracy of EOV data collection in relation to location data from animal-born tags. Researchers will have the ability to synchronize ARGOS and LoRaWan-linked animal-born tags with the constellation SDRs. Therefore, the satellites will be able to collect location data from tagged animals within the FOV. Additionally, transmissions from tagged animals will prompt the HSI to take an image, providing researchers access to near real-time oceanographic data.

Ocean color imagery from AEROS' HSI will be analyzed using a deep learning model to detect ocean fronts. The model will be trained with a dataset generated with Landsat images, labeled to indicate the presence of fronts, enabling front identification in ocean color images. The advantages over traditional gradient based methods include less noise, more clearly defined fronts and the confirmation of the presence or absence of a front in each image. The model will allow end users to have access to fast and accurate information on frontal zones of interest, such as the Azores front. AEROS will streamline the data acquisition process by collecting environmental and animal-tracking data that will be processed at the DAC (Fig.15).

### 4.2. Fisheries and Aquaculture Management

AEROS will deliver data to monitor water quality and primary production in relation to fisheries hotspots (i.e., areas of high space-use) and aquaculture. Ocean color products, such as chlorophyll α concentration (Chl-a), total suspended matter (TSM), and Colored Dissolved Organic Matter (CDOM) will be derived from the AEROS HSI. Combined analyses of these datasets with other environmental parameters (i.e., SST, ocean currents, wave height, SSS, underwater irradiance) from large missions and oceanic numerical models will support sustainable use of marine resources, improved fish health, improved waste management practices, and will be used to monitor/predict Harmful Algae Blooms (HABs). The DAC will combine these products as a more complete tool to end-users. Additionally, these datasets can be overlaid with fishing vessel locations, acquired from vessels equipped with an Automatic Identification System and/or a Vessel Monitoring System, to monitor fishery activities. Lastly, AEROS will provide a set of ocean variables in higher resolution than currently available datasets, which could foster innovative approaches to improve knowledge in small scale processes. This is paramount for design and implementation of fisheries legislation, strengthens the basis for fisheries management and supports efficient aquaculture practices.

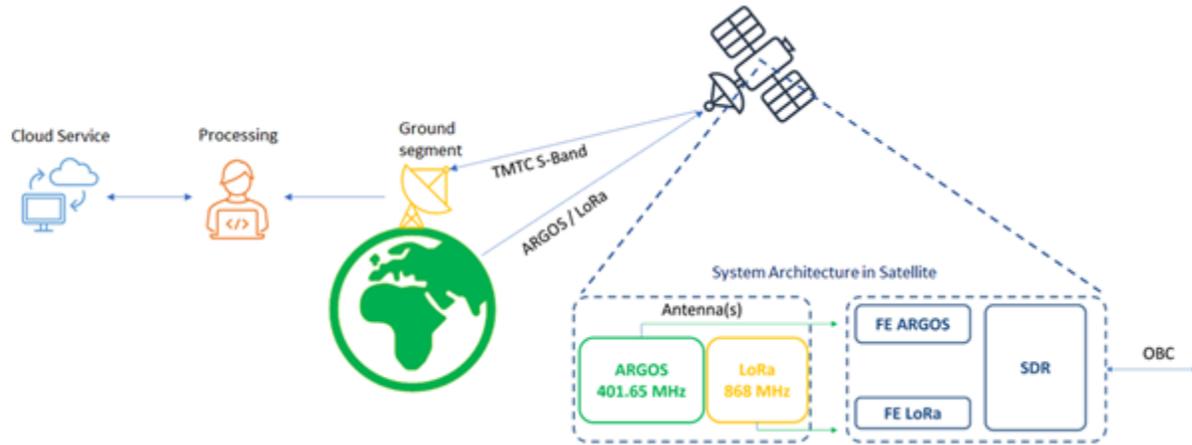

Fig. 15. Schematic illustrating the collection, processing, and transfer of data from remote platforms and onboard sensors (HSI, RGB Camera, and SDR) to end users

*4.3. Monitoring Ecosystem Health and Marine Protected Areas*

Despite increasing efforts to conserve ecosystems, global analyses show alarming rates of loss and change at all levels of biological diversification. As a response to prevent the further degradation and destruction of ecosystem health, the United Nations declared 2021-2030 as the Decade on Ecosystem Restoration [14].

Coastal and marine ecosystems are rich in biodiversity and support a range of purposes including coastal protection, food, water filtration, carbon sink, recreation areas, and cultural value. These complex ecosystems have experienced anthropogenic impacts and the assessment of their good condition can be challenging depending on indicators such as: pressures, physical parameters, biological structures, functional structures, and ecological models. The effectiveness of the chosen indicators can be affected by lack of data and limited research capacity, which in turn changes the outcome of ecosystem condition assessment [15].

MPAs are designated sections of the ocean, coming in a variety of forms and shapes, that restrict anthropogenic activities (such as fishing) to a certain degree. These areas are based on a wide range of legislative instruments and have been used worldwide as central tools for marine conservation, helping to tackle over-exploitation of resources and ecological damage [16].

Areas where MPAs have been implemented have been found to experience positive ecological effects such as increasing species abundance, socio-economic benefits and preventing endangered habitats from experiencing continued decline [17]. However, global MPAs coverage is still less than 10% of the ocean surface (Fig. 16), with no-take MPAs (marine reserves) covering less than 1% of the global ocean. The 30x30 MPAs target, strongly supported by the European Commission's Green Deal, intends to achieve 30% MPAs coverage with 10% strictly protected, by 2030.

MPAs around the world cover a range of different ecosystems, located in different climate conditions and subject to diverse environmental phenomena. Moreover, MPAs have different objectives, such as recovery of a fish stock or conservation of an area. These factors demand different variables to be monitored according to the specific characteristics of each MPA.

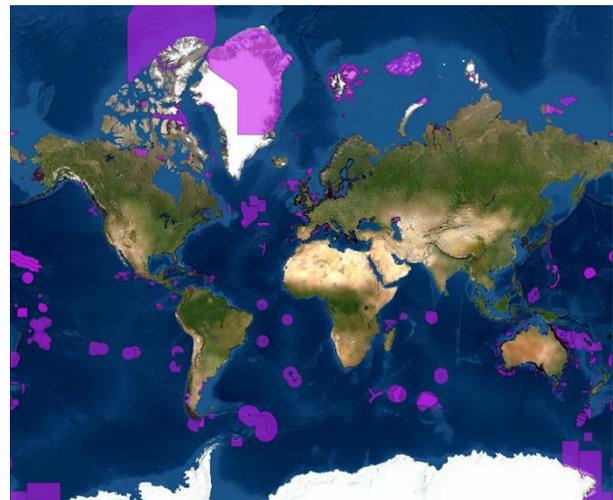

Fig. 16. World distribution of MPAs according to the World Database on Protected Areas (WDPA). Image by Colab+ Atlantic using MPA shapefiles from WDPA.

These necessities can be more efficiently covered by constellations of satellites equipped with different sensors, and that allows for more complete spatial and temporal coverage. As an example of phenomena which should be monitored in MPAs, chemical pollution can be related to different substances, but in the context of coastal MPAs, the most common problems are related to residues from agricultural and urban areas that reach the ocean. These residues are usually composed of nutrients, such as phosphate and nitrate, which can cause eutrophication and, consequently, phytoplankton blooms and anoxic zones. The AEROS HSI retrieved reflectance can be used for nutrient estimation [18], monitoring concentration in coastal waters and alerting of possible eutrophication conditions. Moreover, the 55 m resolution of the HSI is sufficient for this type of analysis [19].

As MPAs are continuously being declared around the globe, not everything is positive and tensions between the conservation and resource exploitation sector are rising. This is in addition to key threats such as eutrophication, plastic pollution, oil spills, climate change, and the risk of invasive species that affect MPAs and their management [16].

These threats can have an effect even if they are happening externally, and thus a constant data feed from satellites is key to trigger and pinpoint in-situ data acquisitions inside MPAs, aiming at monitoring and evaluating their effects. AEROS will be capable of delivering important information on physical and biochemical MPA indicators. It can do this by providing information with the HSI that is useful for distinguishing different phytoplankton classes or species, which have different spectral signatures, enabling the detection of HABs [20]. Besides that, HSI data can also be used to retrieve phytoplankton absorption spectra and biomass with smaller errors than ocean color algorithms [21]. Combined use of SSS and SST data from other sources, enables researchers to study the variability of the thermohaline circulation through ocean-atmosphere heat exchange and water density changes, which may affect ocean MPAs. Collection of CDOM data allow observation of the impacts from terrestrial pollution in coastal MPAs [22]. The previous chemical and physical parameters are recommended for measurement in MPAs, as they provide a direct or indirect indication of the stability of marine communities [23].

## 5. Conclusion/Summary

A successful demonstration of the AEROS mission will serve as a precursor to a future ocean-sensing CubeSat constellation. This constellation will provide global observations of critical oceanographic features with significantly higher spectral, spatial, and temporal resolution. Additionally, this mission serves to space-qualify the Spin.Works' HSI configuration, CEiiA's structural and system integration design, as well as the RBG imager. Last, use of the SDR as a gateway for LoRa demonstrates a novel use case for integrating a CubeSat-based communications payload in the LoRaWAN network.

A successful demonstration of AEROS' HSI will provide high resolution images of the Portuguese coast to survey key oceanographic parameters. Collected data products have the potential to drive innovative approaches for analyzing small-scale oceanographic processes, monitoring MPAs and informing sustainable fishery management. Successful implementation of the Data Analysis Center will provide a tool for end users to easily access the raw and pre-processed data for a gamut of analysis types.

HSI development is ongoing. Though spatial corrections are now well-characterized, next steps involve testing to account for necessary spectral corrections. Additionally, it is intended to transfer the pushbroom simulator outdoors to survey the HSI's performance in broad daylight using a Macbeth Colorchecker chart, a tool used for color calibration with well spectrally characterized samples.

## Acknowledgements

AEROS is being developed by a consortium of 11 Portuguese companies, research institutes and universities, as well as MIT, and is intended to develop Portuguese technologies and competencies to monitor and quantify the benefits of the ocean. CEiiA is responsible for satellite integration and launch, development of the satellite structure, and the DAC to store data products produced by AEROS and associated ground-based sensors. Spin.Works is developing the HSI. DSTelecom will build the SDR for Argos and LoRaWan communication. UMinho will develop Artificial Intelligence models for ocean simulation and detection of mesoscale fronts/eddies. FCUP is also supporting the development of the DAC and the assembly of the satellite. IST is supporting the business case development. Colab+Atlantic is responsible for the use case definitions with the support of UAlgarve, AIR Centre and IMAR. MIT is providing systems engineering and orbital dynamics support. EDISOFT is leading the development of the on-board software and the ground segment. We acknowledge AGI, an Ansys company, and its Educational Alliance Program for donating its Systems Tool Kit software, which was used to conduct analysis for the AEROS mission.

The AEROS project (reference POCI-01-0247-FEDER-045911) leading to this work is co-financed by the ERDF - European Regional Development Fund through the Competitiveness and Internationalisation - COMPETE 2020, PORTUGAL 2020, LISBOA 2020 and by the Portuguese Foundation for Science and Technology - FCT under the International Partnership programme MIT Portugal. The Azorean partners are co-financed by AÇORES 2020, project number ACORES-01-0145-FEDER-000131, by FCT under the project

UIDB/05634/2020, DRCT the Regional Government of the Azores through the initiative to support the Research Centres of the University of the Azores and through the project M1.1.A/REEQ.CIENTÍFICO UI&D/2021/010.

## References

[1] le Quéré, C., Moriarty, R., Andrew, R. M., Canadell, J. G., Sitch, S., Korsbakken, J. I., Friedlingstein, P., Peters, G. P., Andres, R. J., Boden, T. A., Houghton, R. A., House, J. I., Keeling, R. F., Tans, P., Arneth, A., Bakker, D. C. E., Barbero, L., Bopp, L., Chang, J., … Zeng, N. Global Carbon Budget 2015. Earth System Science Data. 7:2 (2015) 349-396.

[2] Cheng, L., Trenberth, K. E., Fasullo, J., Boyer, T., Abraham, J., & Zhu, J. Improved estimates of ocean heat content from 1960 to 2015. In Science Advances. 3:3 (2017) e1601545.

[3] Levitus, S., Antonov, J. I., Boyer, T. P., Baranova, O. K., Garcia, H. E., Locarnini, R. A., Mishonov, A. v., Reagan, J. R., Seidov, D., Yarosh, E. S., & Zweng, M. M. World ocean heat content and thermosteric sea level change (0-2000m), 1955-2010. Geophysical Research Letters. 39:10 (2012) L10603.

[4] von Schuckmann, K., le Traon, P. Y., Alvarez-Fanjul, E., Axell, L., Balmaseda, M., Breivik, L. A., Brewin, R. J. W., Bricaud, C., Drevillon, M., Drillet, Y., Dubois, C., Embury, O., Etienne, H., Sotillo, M. G., Garric, G., Gasparin, F., Gutknecht, E., Guinehut, S., Hernandez, F., … Verbrugge, N. The Copernicus Marine Environment Monitoring Service Ocean State Report. Journal of Operational Oceanography. 9:2 (2016) s235-s320. https://doi.org/10.1080/1755876X.2016.1273446

[5] Field, C. B., Barros, V. R., Dokken, D. J., Mach, K. J., Mastrandrea, M. D., Bilir, T. E., Chatterjee, M., Ebi, K. L., Estrada, Y. O., Genova, R. C., Girma, B., Kissel, E. S., Levy, A. N., MacCracken, S., Mastrandrea, P. R., & White, L. L. (2014). Climate change 2014 impacts, adaptation and vulnerability: Part A: Global and sectoral aspects: Working group II contribution to the fifth assessment report of the intergovernmental panel on climate change. In Climate Change 2014 Impacts, Adaptation and Vulnerability: Part A: Global and Sectoral Aspects. Cambridge University Press. (2014) 1132.

[6] Miloslavich, P, Bax, NJ, Simmons, SE, et al. Essential ocean variables for global sustained observations of biodiversity and ecosystem changes. Glob Change Biol. 24 (2018) 2416–2433.

[7] Toorian, A., Diaz, K., & Lee, S. (2008). The CubeSat approach to space access, 1095-323X, IEEE Aerospace Conference Proceedings, Big Sky, MT, USA, 2008, 01 - 08 March.

[8] Sweeting, M. N. Modern Small Satellites-Changing the Economics of Space. Proceedings of the IEEE. 106:3 (2018) 343-361.

[9] Payne, C., Lifson, M., Thieu, A., Tomio, H., Siew, P. M., Haughwout, C., Newman, D., Cahoy, K., Santos, A., Hormigo, T., Coutinho, M., & Silva, H. The aeros mission: Multi-spectral ocean science measurement network via small satellite connectivity, ASCEND, Las Vegas, Nevada, USA, 2021, 15 – 17 November.

[10] Guerra, A. G., Francisco, F., Villate, J., Agelet, F. A., Bertolami, O., & Rajan, K. On small satellites for oceanography: A survey. Acta Astronautica. 127 (2016) 404-423.

[11] Hussey, Nigel E., Steven T. Kessel, Kim Aarestrup, Steven J. Cooke, Paul D. Cowley, Aaron T. Fisk, Robert G. Harcourt et al. Aquatic animal telemetry: a panoramic window into the underwater world. Science. 348:6240 (2015) 1255642.

[12] Vedor, M., Queiroz, N., Mucientes, G., Couto, A., da Costa, I., Dos Santos, A., ... & Sims, D. W. Climate-driven deoxygenation elevates fishing vulnerability for the ocean's widest ranging shark. Elife.10 (2021) e62508.

[13] Queiroz, Nuno, Nicolas E. Humphries, Ana Couto, Marisa Vedor, Ivo Da Costa, Ana MM Sequeira, Gonzalo Mucientes et al. Global spatial risk assessment of sharks under the footprint of fisheries. Nature. 572:7770 (2019) 461-466.

[14] Aronson, J., Goodwin, N., Orlando, L., Eisenberg, C. and Cross, A.T. A world of possibilities: six restoration strategies to support the United Nation's Decade on Ecosystem Restoration. Restor Ecol. 28 (2020) 730-736.

[15] Smit KP, Bernard AT, Lombard AT, Sink KJ. Assessing marine ecosystem condition: A review to support indicator choice and framework development. Ecological Indicators. 121 (2021) 107148.

[16] Kriegl, Michael, Xochitl E. Elías Ilosvay, Christian von Dorrien, and Daniel Oesterwind. Marine protected areas: at the crossroads of nature conservation and fisheries management. Frontiers in Marine Science. (2021) 627.

[17] Gallacher, J., Simmonds, N., Fellowes, H., Brown N., Gill N., Clark W., Biggs C. and Rodwell L.D. Evaluating the success of a marine protected area: A systematic review approach. Journal of environmental management. 202 (2016) 348-362.

[18] Wang, D., Cui, Q., Gong, F., Wang, L., He, X., & Bai, Y. Satellite retrieval of surface water nutrients in the coastal regions of the East China Sea. Remote Sensing. 10:12 (2018) 1896.

[19] Muller-Karger, F. E., Hestir, E., Ade, C., Turpie, K., Roberts, D. A., Siegel, D., ... & Jetz, W. Satellite sensor requirements for monitoring essential biodiversity variables of coastal ecosystems. Ecological applications. 28:3 (2018) 749-760.

[20] Millie, D. F., Schofield, O. M., Kirkpatrick, G. J., Johnsen, G., Tester, P. A., & Vinyard, B. T. (1997). Detection of harmful algal blooms using

photopigments and absorption signatures: A case study of the Florida red tide dinoflagellate, Gymnodinium breve. Limnology and oceanography. 42 (1997) 1240-1251.

[21] Pahlevan, N., Smith, B., Binding, C., Gurlin, D., Li, L., Bresciani, M., & Giardino, C.Hyperspectral retrievals of phytoplankton absorption and chlorophyll-a in inland and nearshore coastal waters. Remote Sensing of Environment. 253 (2021) 112200.

[22] Keller, S. et al., Hyperspectral Data and Machine Learning for Estimating CDOM, Chlorophyll a, Diatoms, Green Algae and Turbidity, Int. J. Environ. Res. Public Heal. 15 (2018) 1881.

[23] del Mar Otero M, Garrabou J, Vargas M. Mediterranean Marine Protected Areas and climate change: A guide to regional monitoring and adaptation opportunities. IUCN; 2013.